\documentclass[twocolumn,letterpaper,11pt]{article}
\usepackage[nocopyright,11pt]{sigminplusplus}

\usepackage{subfigure}
\usepackage{verbatim}  
\usepackage{graphicx}
\IfFileExists{url.sty}{\usepackage{url}}
                      {\newcommand{\url}{\texttt}}

\usepackage{bibspacing}
\usepackage{tweaklist} 

\renewcommand{\itemhook}{\setlength{\itemsep}{0pt}}
\renewcommand{\enumhook}{\setlength{\itemsep}{0pt}}

\newcommand{\urlsamefont}[1]
{
\urlstyle{same}\url{#1}
}

\makeatletter

\usepackage{cite}
\makeatother

\begin{document}

\title{\textbf{Small Is Not Always Beautiful\footnote{Appeared in IPTPS'2008, Tampa Bay, Florida, USA.}}}

\numberofauthors{4}
\author{
\alignauthor 
  \authorname{Pawe{\l} Marciniak\thanks{Work done while an 
intern at INRIA Sophia Antipolis.}}\\
  \affinfo{Poznan University of Technology, Poland}\\
  \email{pawel.marciniak@gmail.com}
\alignauthor 
  \authorname{Nikitas Liogkas}\\
  \affinfo{UCLA}\\
  \affinfo{Los Angeles, CA}\\
  \email{nikitas@cs.ucla.edu}
\alignauthor 
  \authorname{Arnaud Legout}\\
  \affinfo{I.N.R.I.A.}\\
  \affinfo{Sophia Antipolis, France}\\
  \email{arnaud.legout@sophia.inria.fr}
\alignauthor 
  \authorname{Eddie Kohler}\\
  \affinfo{UCLA}\\
  \affinfo{Los Angeles, CA}\\
  \email{kohler@cs.ucla.edu}
}

\copyrightnotice{IPTPS'2008}
\date{}  
\maketitle

\pagestyle{empty}

\begin{abstract}

Peer-to-peer content distribution systems have been enjoying great popularity,
and are now gaining momentum as a means of disseminating video streams 
over the Internet. 
In many of these protocols, including the popular BitTorrent, content is 
split into mostly fixed-size pieces, allowing 
a client to download data from many peers simultaneously.
This makes \textit{piece size} potentially critical for performance.
However, previous research efforts have largely overlooked this parameter, opting 
to focus on others instead.

This paper presents the results of real experiments with varying piece 
sizes on a controlled BitTorrent testbed.
We demonstrate that this parameter is indeed critical, as it determines the degree 
of parallelism in the system, and we investigate optimal piece sizes for
distributing small and large content. We also pinpoint a related design trade-off,
and explain how BitTorrent's choice of dividing pieces into subpieces 
attempts to address it.

\end{abstract}

\section{Introduction}

Implementation variations and parameter settings can severely affect the
service observed by the clients of a peer-to-peer system. 
A better understanding of protocol parameters is
needed to improve and stabilize service, a particularly
important goal for emerging peer-to-peer applications such as streaming video.

BitTorrent is widely regarded as one of the most successful swarming protocols,
which divide the content to be distributed into distinct pieces and enable peers 
to share these pieces efficiently. Previous research efforts
have focused on the algorithms 
believed to be the major factors behind BitTorrent's good performance, such as the 
piece and peer selection strategies. 
However, to the best of our knowledge, no studies have looked into the optimal 
size of content pieces being exchanged among peers.
This paper investigates this parameter by running real experiments with 
varying piece sizes on a controlled testbed, and demonstrates that \emph{piece size 
is critical for performance}, as it determines the degree of parallelism available 
in the system.
Our results also show that, for small-sized content, smaller pieces enable
shorter download times, and as a result, \emph{BitTorrent's design choice of further 
dividing content pieces into subpieces is unnecessary for such content}.
We evaluate the overhead that small pieces incur as content size grows and 
demonstrate a trade-off between piece size and available 
parallelism. We also explain how this trade-off motivates
the use of both pieces and subpieces for distributing large content, the common case 
in BitTorrent swarms.

The rest of this paper is organized as follows. Section~\ref{sec:methodology} 
provides a brief description of the BitTorrent protocol, and describes our 
experimental methodology. Section~\ref{sec:results} then presents the results 
of our experiments with varying piece sizes, while Section~\ref{sec:discussion} 
discusses potential reasons behind the poor performance of small pieces when distributing 
large content.
Lastly, Section~\ref{sec:related} describes related work and 
Section~\ref{sec:conclusion} concludes.

\section{Background and Methodology}
\label{sec:methodology}

\paragraph{BitTorrent Overview}

BitTorrent is a popular peer-to-peer content distribution protocol that has been shown 
to scale well with the number of participating clients. Prior to distribution, the
content is divided into multiple \textit{pieces}, while each piece is further 
divided into multiple \textit{subpieces}. A \textit{metainfo file} containing 
information necessary for initiating the download process is then created by the content 
provider. This information includes each piece's SHA-1 hash (used to verify
received data) and the address of the \textit{tracker}, 
a centralized component that facilitates peer discovery.

In order to join a \textit{torrent}---the collection of peers participating in the
download of a particular content---a client retrieves the metainfo file out of band, 
usually from a Web site. It then contacts the tracker, which responds 
with a \textit{peer set} of randomly selected peers. These might include both 
\textit{seeds}, who already have the entire content and are sharing it with others, 
and \textit{leechers}, who are still in the process of downloading.
The new client can then start contacting peers in this set and request data.
Most clients nowadays implement a \textit{rarest-first} policy for piece requests: 
they first ask for the pieces that exist at the smallest number of peers in their 
peer set. 
Although peers always exchange just subpieces with each other, they only
make data available in the form of complete pieces: after downloading
all subpieces of a piece, a peer notifies all peers in its peer set with a 
\textit{have} message.
Peers are also able to determine which pieces others have based on a 
\textit{bitfield} message, exchanged upon the establishment of new connections, which
contains a bitmap denoting piece possession.

Each leecher independently decides who to exchange data with via the
\textit{choking algorithm}, which gives preference to those who upload data to 
the given leecher at the highest rates. Thus, once per \textit{rechoke period}, 
typically every ten seconds, a leecher considers the receiving data rates from 
all leechers in its peer set. It then picks out the fastest ones, a fixed number of 
them, and only uploads to
those for the duration of the period. 
Seeds, who do not need to download any pieces, follow a different unchoke strategy. 
Most current implementations unchoke those leechers that \textit{download} data 
at the highest rates, to better utilize seed capacity.

\paragraph{Experimental Methodology}

\begin{figure*}[t]
\centering
\subfigure[Piece size of 16 kB]{\includegraphics[width=0.23\textwidth]
{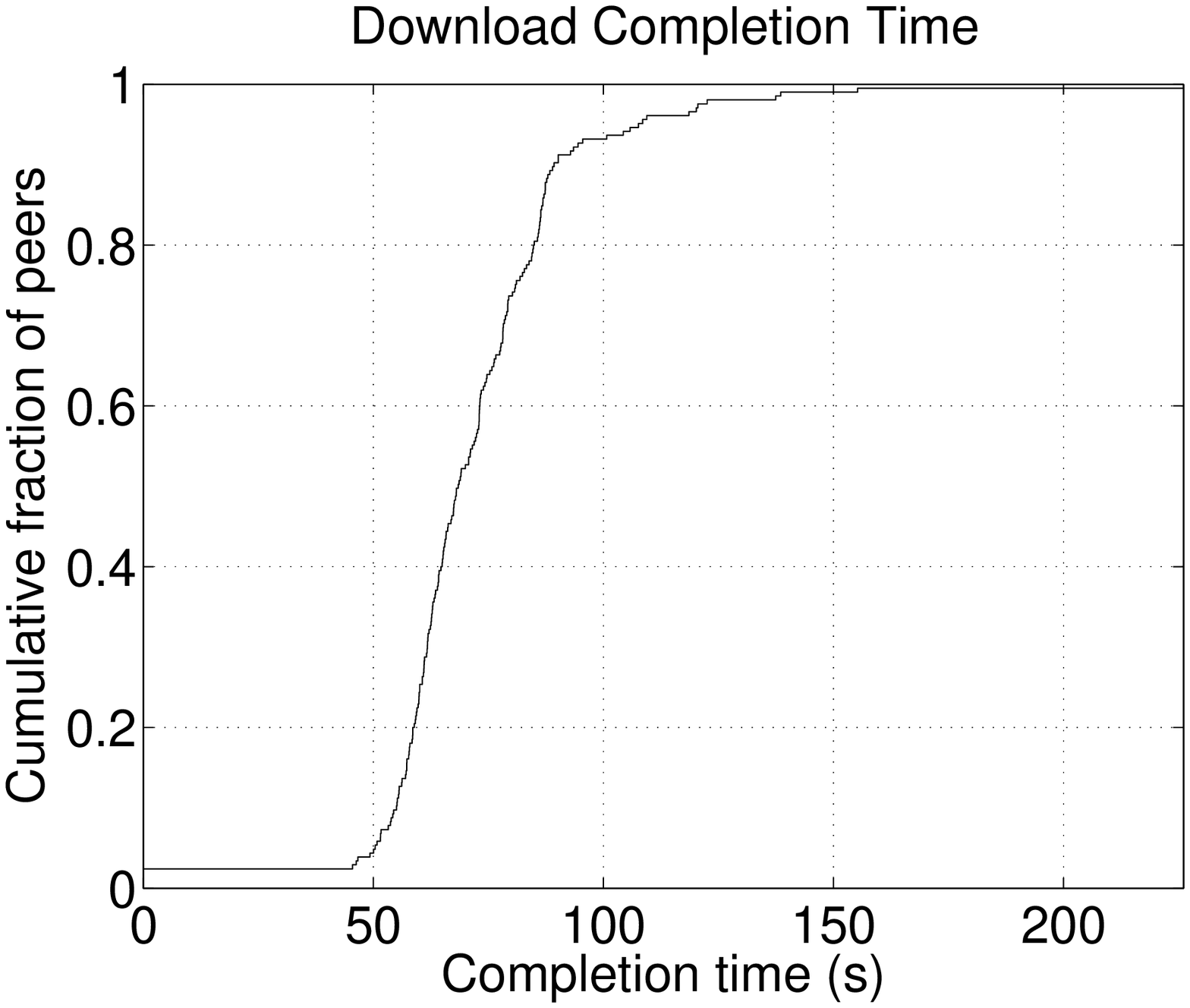}}
\hfill{}
\subfigure[Piece size of 512 kB]{\includegraphics[width=0.23\textwidth]
{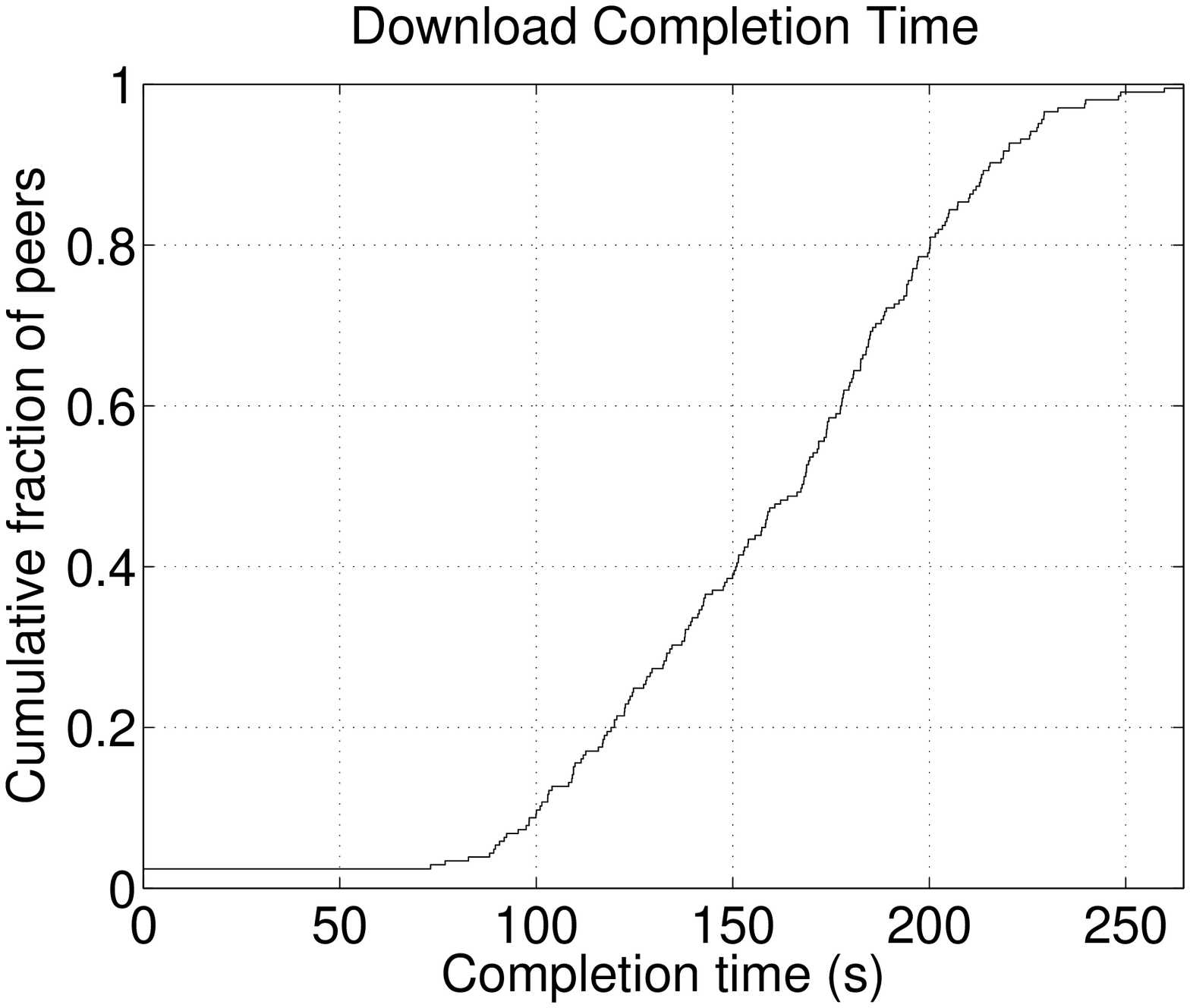}}
\hfill{}
\subfigure[Piece size of 16 kB]{\includegraphics[width=0.23\textwidth]
{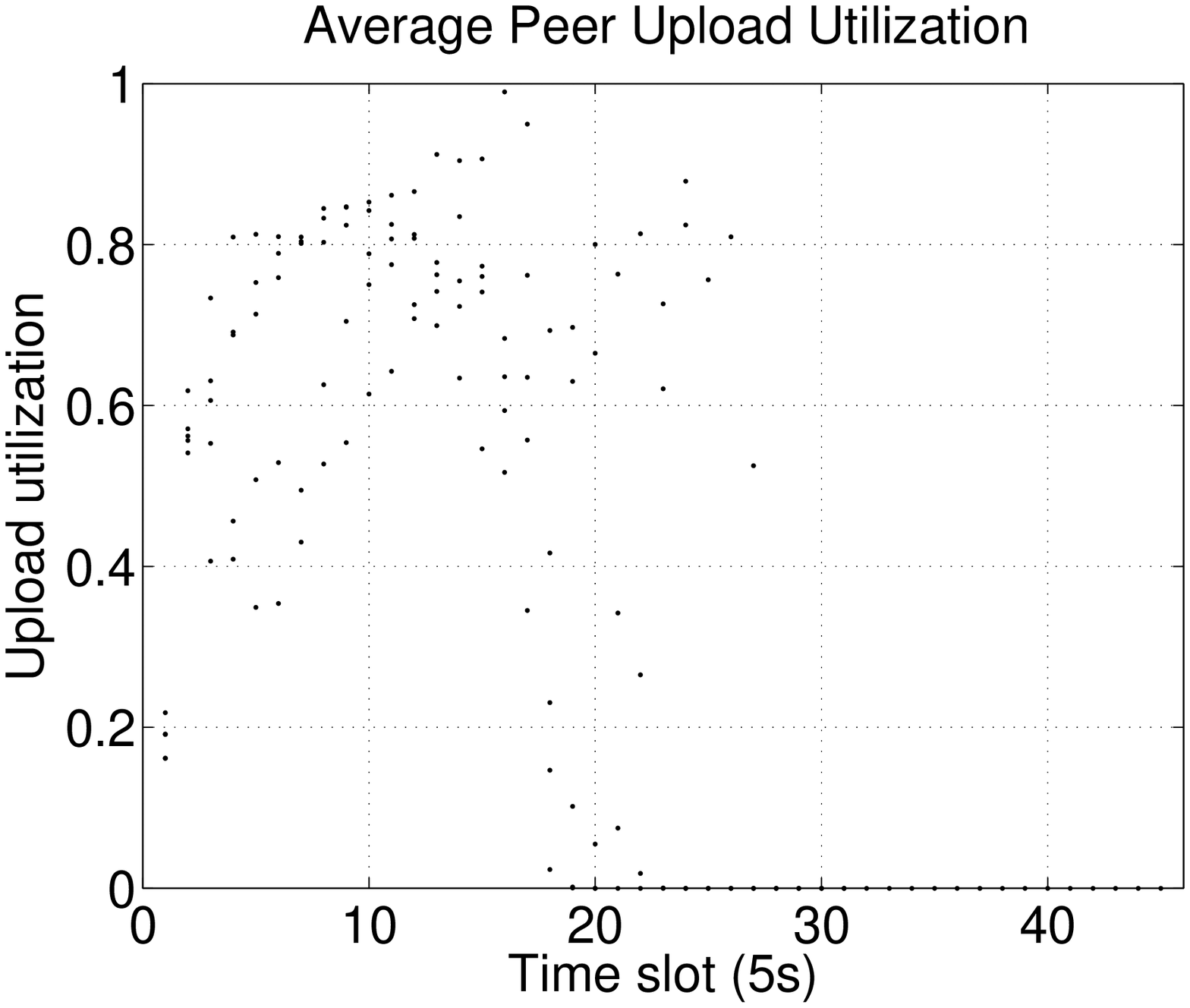}}
\hfill{}
\subfigure[Piece size of 512 kB]{\includegraphics[width=0.23\textwidth]
{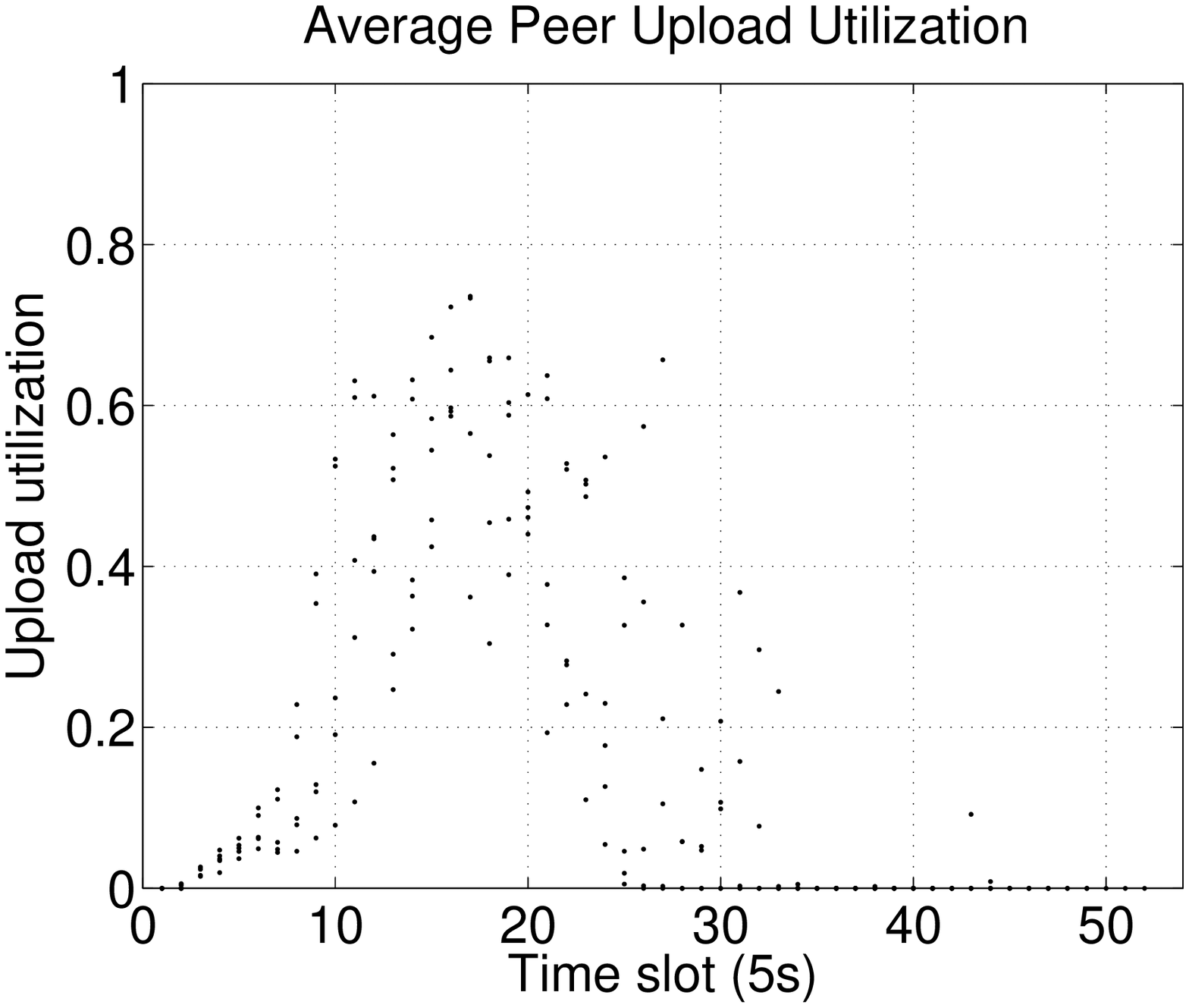}}
\caption{CDFs of peer download completion times and scatterplots of average
upload utilization for five-second time intervals when
distributing a 5~MB content (averages over 5 runs).
\emph{Small pieces shorten download time and enable higher utilization.}}
\label{fig:small_contents-cdfs_utilization}
\end{figure*}

We have performed all our experiments with private torrents on the PlanetLab 
platform~\cite{planetlab}. 
These torrents comprise 40 leechers and a single initial seed sharing content of 
different sizes. Leechers do not change their available upload bandwidth during the 
download, and disconnect after receiving a complete copy of the content. 
The initial seed stays connected for the duration of the experiment, while all leechers 
join the torrent at the same time, emulating a flash crowd scenario. 
The number of parallel upload slots is set to 4 for the leechers and seed.
Although system behavior might be different with other peer arrival patterns and torrent 
configurations, there is no reason to believe that the conclusions we draw are predicated 
on these parameters.

The available bandwidth of most PlanetLab nodes is relatively high for typical real-world clients.
We impose upload limits on the leechers and seed to model more realistic scenarios,
but do not impose any download limits, as we wish to observe differences in download completion time
with varying piece sizes. 
The upload limits for leechers follow
a uniform distribution from 20 to 200~kB/s, while the seed's upload capacity is 
set to 200~kB/s.

We collect our measurements using the official (mainline) BitTorrent 
implementation, instrumented to record interesting events.
Our client is based on version 4.0.2 of the official implementation
and is publicly available for download~\cite{btinstrumented}.
We log the client's internal state, as well as each message sent or received along with the 
content of the 
message. Unless otherwise specified, we run our 
experiments with the default parameters.

The protocol does not strictly define the piece and subpiece sizes.
An unofficial BitTorrent specification~\cite{btwikispec} states that the 
conventional wisdom is to ``pick the smallest piece size that results in a metainfo file 
no greater than 50--75~kB''. The most common piece size for public torrents seems to be 256~kB.
Additionally, most implementations nowadays use 16~kB subpieces.
For our experiments, we always keep the subpiece size constant at 16~kB, and
only vary the piece size. We have results for all possible 
combinations of different content sizes (1~MB, 5~MB, 10~MB, 20~MB, 50~MB, and 100~MB) 
and piece sizes 
(16~kB, 32~kB, 64~kB, 128~kB, 256~kB, 512~kB, 1024~kB, and 2048~kB).

\section{Results}
\label{sec:results}

Our results, presented in this section, demonstrate that small pieces are preferable for the 
distribution of
small-sized content. We also discuss the benefits and drawbacks of small
pieces for other content sizes, and evaluate the communication and
metainfo file overhead that different piece sizes incur for larger content.

\subsection{Small Content}

\begin{figure}
\centering
\includegraphics[width=0.6\columnwidth]{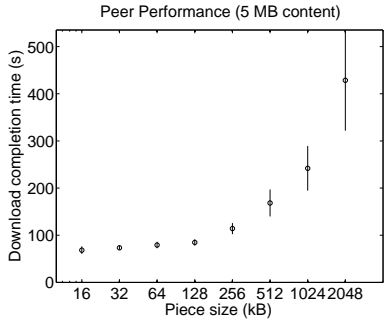}
\caption{Download completion times for a 5~MB content (medians over 5 runs and standard 
deviation error bars). 
\emph{Smaller pieces clearly improve performance.}}
\label{fig:small_contents-completion_comparison}
\end{figure}

Even though most content distributed with BitTorrent is large, it is still interesting to examine the impact
of piece size on distributing smaller content. In addition to gaining a better understanding of the 
trade-offs involved, it may also sometimes be desirable to utilize BitTorrent to avoid server overload when 
distributing small content, e.g., in the case of websites that suddenly become popular.
Figure~\ref{fig:small_contents-completion_comparison} shows the median download 
completion times of the 40 leechers downloading a 5~MB file, for
different numbers of pieces, along with standard deviation error bars.
Clearly, \emph{smaller piece sizes enable faster downloads}.
In particular, performance deteriorates rapidly when increasing the piece size
beyond 256~kB.
The same observations hold for experiments with other small content
(1 and 10~MB).

To better illustrate the benefits of small pieces,
Figure~\ref{fig:small_contents-cdfs_utilization}
shows the cumulative distribution functions (CDF) of leecher download 
completion times for 16 and 512~kB pieces (graphs (a) and (b)).
With small pieces, most peers complete their download within the first 100 seconds.
With larger pieces, on the other hand, the median peer completes in more than twice the 
time, and there is greater variability.
The reason is that \emph{smaller pieces let peers share data sooner.} 
As mentioned before,
peers send out \textit{have} messages announcing new 
pieces only after downloading and verifying a complete piece. Decreasing piece 
size allows peers to download complete pieces, and thus start sharing them with others, sooner.
This increases the available parallelism in the system, as it enables more opportunities
for parallel downloading from multiple peers.

This benefit is also evident when considering \textit{peer upload utilization}, 
which constitutes a reliable metric of efficiency, since the total peer upload 
capacity represents the maximum throughput the system can achieve as a whole. 
Figure~\ref{fig:small_contents-cdfs_utilization} shows utilization scatterplots 
for all five-second time intervals during the download (graphs (c) and (d)). 
Average upload utilization for each of 5 experiment runs is plotted once every 5 seconds.
Thus, there are five dots for every time slot, representing the average peer upload utilization 
for that slot in the corresponding run. The metric is torrent-wide: for those five seconds, 
we sum the upload 
bandwidth expended by leechers and divide by the available upload capacity of all 
leechers still connected to the system. 
Thus, a utilization of 1 represents taking full advantage of the available upload capacity.
As previously observed~\cite{legout07}, utilization is low at the beginning and end of the 
session. During the majority
of the download, however, a smaller piece size increases the number of 
pieces peers are interested in, which leads to higher upload utilization.

\begin{figure}[t]
\centering
\begin{tabular}{@{}p{.5\hsize}@{}p{.5\hsize}@{}}
\subfigure[Piece size of 16 kB]{\includegraphics[scale=0.22]
{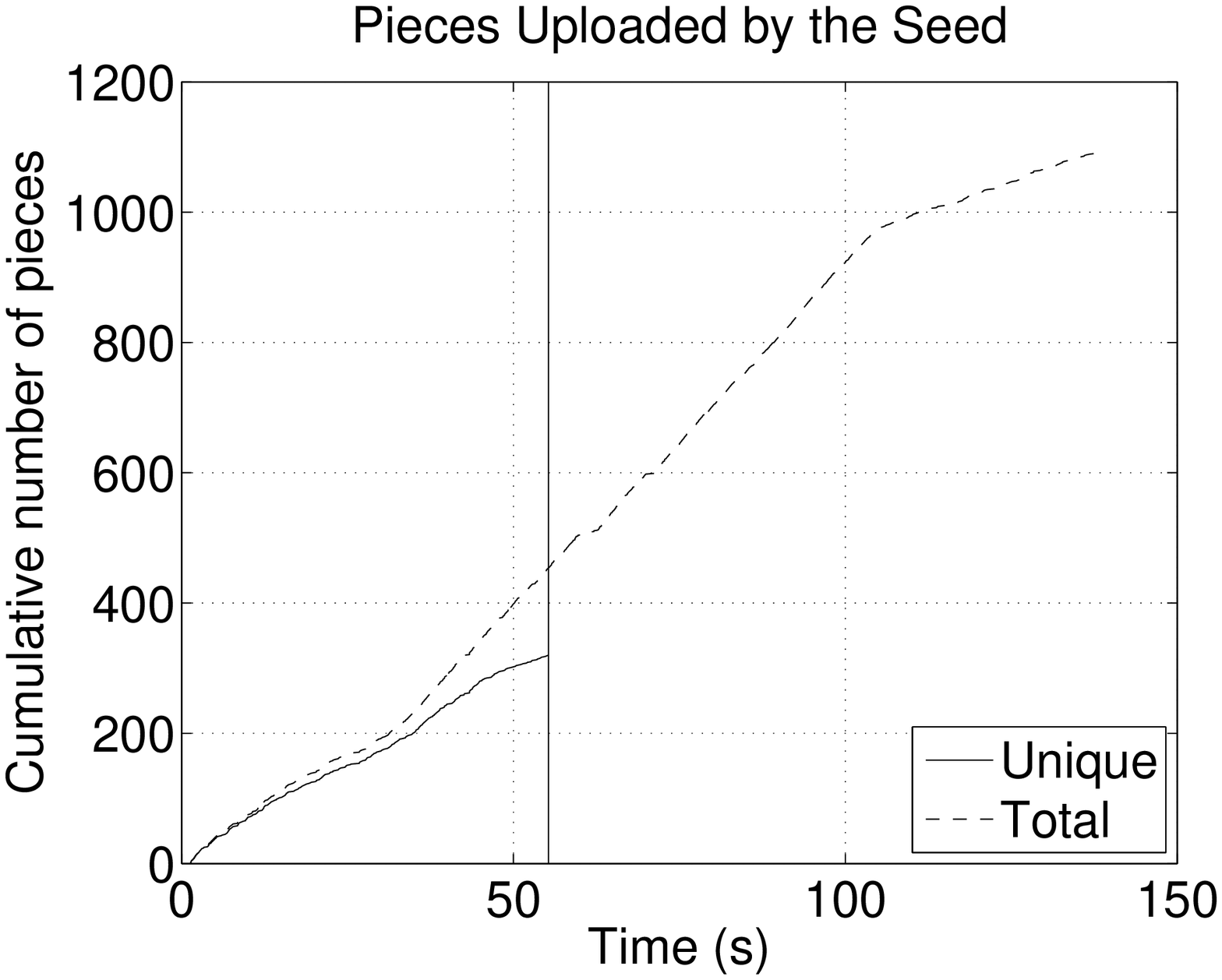}} &
\subfigure[Piece size of 512 kB]{\includegraphics[scale=0.22]
{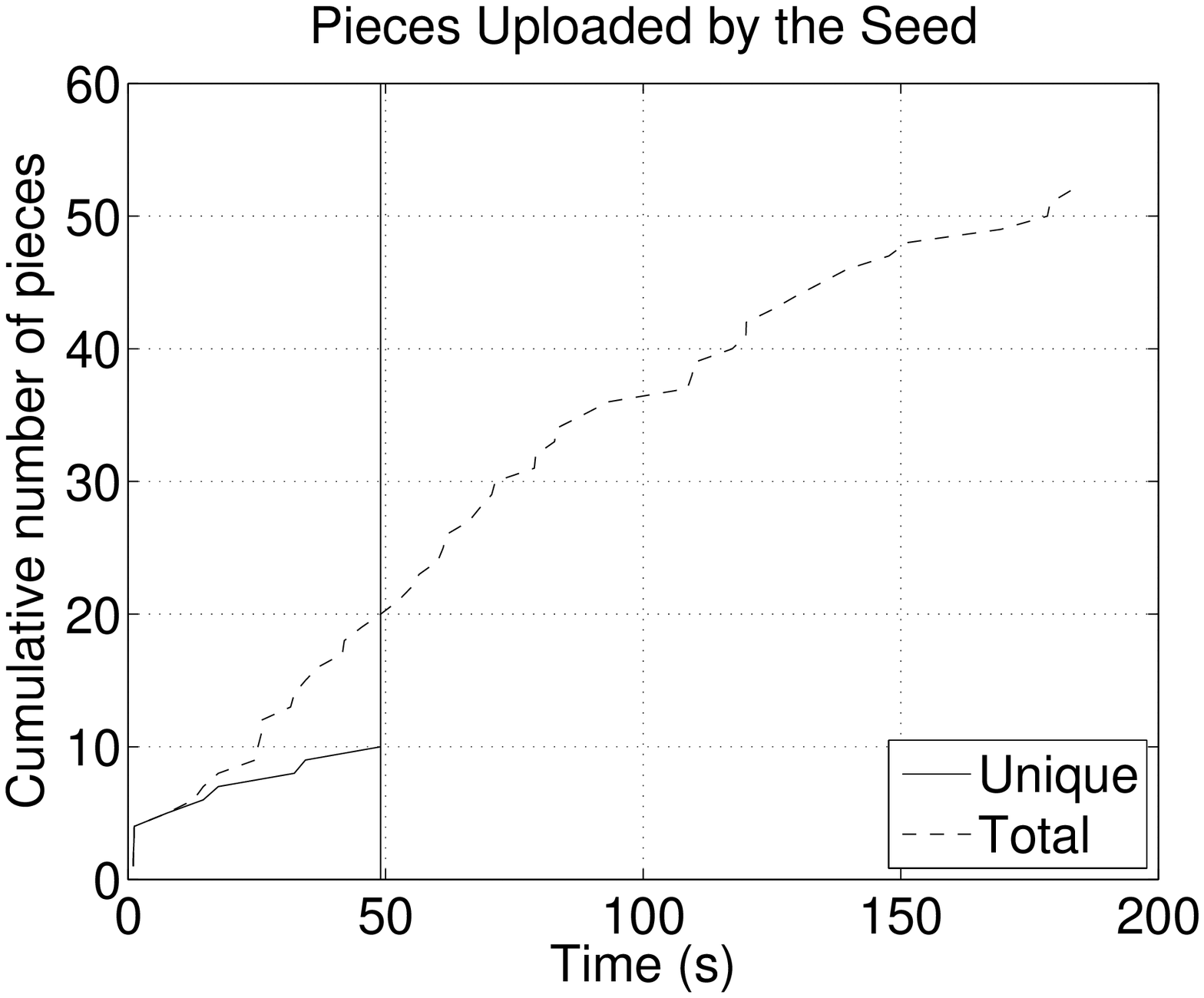}}
\end{tabular}
\caption{Number of pieces uploaded by the seed when distributing a 5~MB content, 
for two representative runs.
The \textit{Unique} line represents the pieces that had not been previously uploaded, 
while the \textit{Total} line represents the total number of pieces uploaded so far.
The vertical line denotes the time the seed finished uploading the first copy of
the content to the system. 
\emph{The duplicate piece overhead is significantly lower for small pieces.}}
\label{fig:small_contents-seed_duplicates}
\end{figure}

These conclusions are reinforced by the fact that small pieces enable the seed to 
upload less duplicate pieces 
during the beginning of a torrent's lifetime.
Figure~\ref{fig:small_contents-seed_duplicates} indeed plots the number of pieces 
(unique and total) uploaded by the single seed in our 5~MB experiments, for two
representative runs. Although the seed finishes uploading the first copy of
the content at approximately the same time in both cases (vertical line on the graphs), 
it uploads 139\% more
duplicate data with larger pieces (5120~kB for 512~kB pieces vs. 2144~kB for 16~kB pieces), 
thus making less efficient use
of its valuable upload bandwidth.
Avoiding this waste can lead to better performance,
especially for low-capacity seeds~\cite{legout07}.
This behavior can be explained as follows. The official BitTorrent implementation
we are using always 
issues requests for the 
rarest pieces \emph{in the same order}. As a result, while a leecher is downloading 
a given piece, other leechers might end up requesting the same piece from the seed. 
With smaller pieces, the time interval before a piece is completely downloaded and shared 
becomes shorter, mitigating this problem. This could be resolved by having 
leechers request rarest pieces in random order instead.

In summary, small pieces enable significantly better performance when distributing small content.
As a result, \emph{the distinction of pieces and subpieces that the BitTorrent design 
dictates is unnecessary for such content}. For instance, in our 5~MB experiments, 
pieces that are as small as subpieces (16~kB) are optimal. Thus, the content 
could just be divided into pieces with no loss of performance.

\subsection{Piece Size Impact}

Before investigating the impact of piece size on the distribution of larger content,
let us first examine the advantages and drawbacks of small pieces.
We have seen that their benefits for small content are largely due 
to the increased peer upload utilization such pieces enable.
Since small pieces can be downloaded sooner than large ones,
leechers are able to share small pieces sooner. In this manner, 
there is more data available in the system, which gives peers a wider
choice of pieces to download. In addition to this increased parallelism, small pieces
provide the following benefits (some of which do not affect our experiments).

\begin{itemize}
\item They decrease the number of duplicate 
pieces uploaded by seeds, thereby better utilizing seed upload
bandwidth.
\item The rarest-first piece selection strategy is more effective in ensuring piece replication.
A greater number of pieces to choose from entails a lower probability that peers
download the same piece, which in turn improves the diversity of pieces
in the system.
\item There is less waste when downloading corrupt data.
Peers can discover bad pieces sooner and re-initiate their download.
\end{itemize}

On the other hand, for larger content, the overhead incurred by small pieces may
hurt performance. This overhead includes the following.

\begin{itemize}
\item Metainfo files become larger, since they have to include more SHA-1 hashes.
This would increase the load on a Web
server serving such files to clients, especially in a flash crowd case.
\item \textit{Bitfield} messages also become larger due to the increased number of bits 
they must contain.
\item Peers must send more \textit{have} messages, resulting in increased
communication overhead.
\end{itemize}

In the next section, we shall see that these drawbacks of small pieces outweigh their benefits, 
for larger content.
Thus, the choice of piece size for a download should take
the content size into account.

\subsection{Larger Content}

\begin{figure}[t]
\centering
\includegraphics[width=0.6\columnwidth]{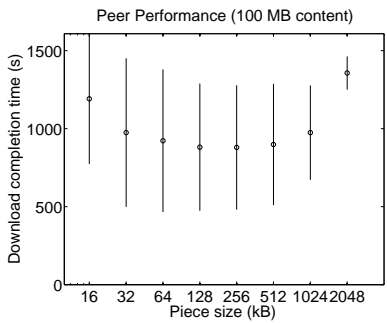}
\caption{Download completion times for a 100~MB content (medians over 5 runs and standard 
deviation error bars).
\emph{Small pieces are no longer optimal.}}
\label{fig:large_contents-completion_comparison}
\end{figure}

Figure~\ref{fig:large_contents-completion_comparison} shows the download 
completion times of the 40 leechers downloading a 100~MB file for different
piece sizes. 
We observe that small pieces are no longer optimal. In this particular
case, sizes around 256~kB seem to perform the best.
Experiments with other content sizes (20 and 50~MB) show that \emph{the optimal piece 
size increases with content size}. 
For instance, for experiments with a 50~MB content, the optimal piece size is 64~kB.
Note that the unofficial guideline for choosing the piece size, mentioned in 
Section~\ref{sec:methodology}, would yield sizes of 32 and 16~kB for a 100~MB and 50~MB
content respectively, a bit off from the optimal values.

\begin{figure}
\centering
\includegraphics[width=0.6\columnwidth]{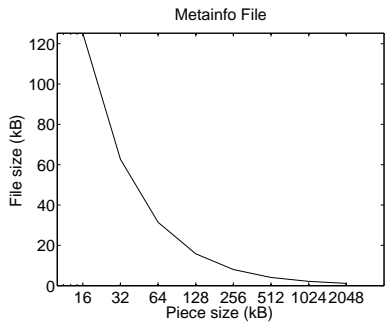}
\caption{Metainfo file sizes for distributing a 100~MB content.
\emph{Smaller pieces produce proportionately larger files.}}
\label{fig:large_contents-metainfo_size}
\end{figure}

In an effort to better understand this trade-off regarding the choice of piece size, we evaluate
the metainfo file and communication overhead. The former is
shown in Figure~\ref{fig:large_contents-metainfo_size}.
As expected, small pieces produce proportionately larger metainfo files (note that
the x axis is logarithmic).
16~kB pieces, for instance, 
produce a metainfo file larger than 120~kB, as compared to a less 
than 10~kB file for 256~kB pieces.
For large content in particular, this might have significant negative implications for 
the Web server used to distribute such files to clients.

\begin{figure}[t]
\centering
\includegraphics[width=0.6\columnwidth]{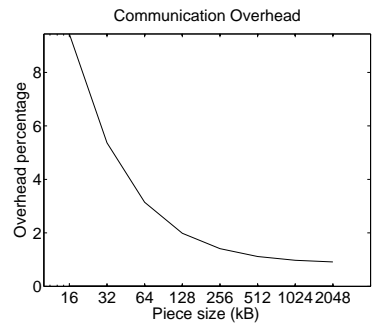}
\caption{Communication overhead due to \textit{bitfield} and \textit{have} messages 
when distributing a 100~MB content.
\emph{Small pieces incur considerably larger overhead.}}
\label{fig:large_contents-overhead}
\end{figure}

\textit{Bitfield} messages become proportionately larger too. For instance, 
for the 100~MB content, these messages are 805 and 55 bytes for 16 and 256~kB pieces 
respectively. Figure~\ref{fig:large_contents-overhead} additionally shows the communication overhead
due to \textit{bitfield} and \textit{have} messages, expressed as a percentage
of the total upload traffic per peer. The overhead ranges from less than 
1\% for larger
piece sizes to around 9\% for 16~kB pieces. However, it is not clear that this overhead
is responsible for the worse performance of smaller pieces.
Although these control messages do occupy upload bandwidth, they do not necessarily 
affect the data exchange among peers, and thus their download performance.
For example, looking at the corresponding overhead for smaller content, we observe that the overhead curve 
looks very similar. This indicates that \emph{increased communication
overhead is 
most likely neither the cause of the worse performance of small pieces for larger
content}, nor does it explain the observed trade-off. 
In the next section, we formulate two hypotheses that might 
help identify the true cause of this behavior.

In summary, when distributing larger content, the optimal 
piece size depends on the content size, due to a trade-off between 
the increased parallelism small pieces provide and their drawbacks.
\emph{BitTorrent arguably attempts to address this trade-off by further
dividing pieces into subpieces}, to get the best of both worlds: 
subpieces increase opportunities for parallel downloading, while pieces 
mitigate the drawbacks of small verifiable units.

\section{Discussion}
\label{sec:discussion}

The results presented in the previous section point to a hidden reason behind
the poor performance of small pieces when distributing large content. We have
two hypotheses that might help explain that.

First, small pieces \emph{reduce opportunities for subpiece request pipelining}.
In order to prevent delays due to request/response latency, and to keep the download
pipe full most of the time, most BitTorrent implementations
issue requests for several consecutive subpieces back-to-back. This 
pipelining, however, is restricted within the boundaries of a single piece.
This is done in order to use available bandwidth to download complete pieces as soon 
as possible, and share them
with the rest of the swarm. Similarly, peers do not typically issue a request for 
subpieces of another piece 
to the same peer before completing the previous one.
Thus, for a content with 32~kB pieces, for instance, only two subpiece requests per peer 
can be pending at any point in time.
For small content, the impact of reduced pipelining is negligible,
as the download completes quickly anyway.
For larger content, however, it might severely affect system performance, as it
limits the total number of simultaneous requests a peer can issue.
Additionally, this matter
gains importance as available peer bandwidth rises, since the request/response
latency then starts to dominate time spent on data transmission.

Furthermore, small pieces \emph{may incur slowdown due to TCP effects}. With a small piece size, a given 
leecher is more likely to keep switching among peers to download different pieces of the content.
This could have two adverse TCP-related effects: 1) the congestion window would have less time 
to ramp up than
in the case of downloading a large piece entirely from a single peer, and 2) the congestion 
window for unused peer connections would gradually decrease after a period 
of inactivity, due to TCP congestion window validation~\cite{rfc2861}, which
is enabled by default in recent Linux kernels, such as the ones running on the PlanetLab 
machines in our experiments.
A large piece size, on the other hand, would enable more efficient TCP transfers due to 
the lower probability 
of switching from peer to peer. Note, however, that, even in that case, there is no guarantee
that all subpieces of a piece will be downloaded from the same peer.

\section{Related Work}
\label{sec:related}

To the best of our knowledge, this is the first study that systematically investigates
the optimal piece size in BitTorrent.
Bram Cohen, the protocol's creator, first described BitTorrent's main 
algorithms and their design rationale~\cite{cohen03}. 
In version 3.1 of the official implementation he reduced the default piece size from 
1~MB to 256~kB~\cite{piecechange}, albeit without giving a concrete reason for 
doing so. Presumably, he noticed the performance benefits of smaller pieces.

Some previous research efforts have looked into the impact of piece size in other 
peer-to-peer content distribution systems.
Ho{\ss}feld \textit{et al.}~\cite{hossfeld05} used simulations
to evaluate varying piece sizes in an eDonkey-based mobile file-sharing system. They found that
download time decreases with piece size up to a certain point,
confirming our observations, although they did not attempt to explain this behavior.
The authors of Dandelion~\cite{sirivianos07} evaluate its performance with different piece sizes, and mention
TCP effects as a potential reason for the poor performance of small pieces. However, small
pieces in that system may also be harmful because they increase the rate at which key requests are 
sent to the central server.
CoBlitz~\cite{park06} faces a similar problem with smaller pieces requiring more processing
at CDN nodes. The authors end up choosing a piece size of 60~kB, because that can easily fit into 
the default Linux outbound kernel socket buffers.
The Slurpie~\cite{sherwood04} authors briefly allude to a piece size trade-off, and 
mention TCP overhead as a drawback of small pieces.
Lastly, during the evaluation of the CREW system~\cite{deshpande06},
the authors find a piece size of 8~kB to be optimal for distributing a
800~kB content, but they do not attempt to explain that.

\section{Conclusion}
\label{sec:conclusion}

This paper presents results of real experiments with varying piece sizes on a 
controlled BitTorrent testbed. We show that piece size is critical for performance, 
as it determines the degree of parallelism in the system.
Our results explain why small pieces are the optimal choice for small-sized content,
and why further dividing content pieces into subpieces is unnecessary for such 
content.
We also evaluated the overhead small pieces incur for larger content, and 
discussed the design trade-off between piece size and available parallelism.

It would be interesting
to investigate our two hypotheses regarding the poor performance of small pieces
with larger content.
We would also like to extend our conclusions to different scenarios, such as 
video streaming, which imposes additional real-time constraints on the protocol.

\paragraph{Acknowledgments} 
We are grateful to Michael Sirivianos, Himabindu Pucha, and the anonymous reviewers 
for their valuable feedback.

\begin{small}

\end{small}

\end{document}